\documentclass[10pt, prd, twocolumn, showpacs, amsmath, amssymb, aps, floats, floatfix, nofootinbib]{revtex4-1}

\usepackage{bm}
\usepackage{array}
\usepackage{graphicx}
\usepackage{subfigure}
\usepackage{multirow}
\usepackage{dcolumn}
\usepackage[]{color}
\usepackage[CJKbookmarks=true,colorlinks,linkcolor=blue,anchorcolor=red,citecolor=green]{hyperref}
\usepackage{cleveref}
\usepackage{multirow}
\usepackage{setspace}

\def\aap{Astron.\ Astrophys.\ }
\def\apj{Astrophys.\ J.\ }
\def\apjl{Astrophys.\ J.\ Lett.\ }

\def\aj{Astron.\ J.\ }
\def\mnras{Mon.\ Not.\ Roy.\ Astron.\ Soc.\ }
\def\physrep{Phys.\ Rept.\ }
\def\prd{Phys.\ Rev.\ D\ }

\def\araa{Annu.\ Rev.\ Astron.\ Astrophys.\ }
\def\jcap{J.\ Cosmol.\ Astropart.\ Phys.\ }

\newcolumntype{p}{D{,}{\pm}{-1}}

\begin{document}
\title{Constraint on the velocity dependent dark matter annihilation cross section from gamma-ray and kinematic observations of ultrafaint dwarf galaxies}

\author{Yi Zhao$^{1,2}$}
\author{Xiao-Jun Bi$^{3,4}$}
\author{Peng-Fei Yin$^{3}$}
\author{Xinmin Zhang$^{1,2,4}$}

\affiliation{
$^1$ Theoretical Physics Division, Institute of High Energy Physics, \\
Chinese Academy of Sciences, Beijing 100049, China \\
$^2$ Theoretical Physics Center for Science Facilities, \\
Chinese Academy of Sciences, Beijing 100049, China \\
$^3$ Key Laboratory of Particle Astrophysics, Institute of High Energy Physics,\\
Chinese Academy of Sciences, Beijing 100049, China\\
$^4$ School of Physical Sciences, University of Chinese Academy of Sciences, Beijing 100049, China
}

\begin{abstract}
Searching for $\gamma$ rays from dwarf spheroidal galaxies (dSphs) is a promising approach
to detect dark matter (DM) due to the high DM densities and low baryon components in dSphs.
The Fermi-LAT observations from dSphs have set stringent constraints
on the velocity independent annihilation cross section. However, the constraints from dSphs may
change in velocity dependent annihilation scenarios because of the different velocity dispersions
in galaxies. In this work, we study how to set constraints on the velocity dependent annihilation
cross section from the combined Fermi-LAT observations of dSphs with the kinematic data. In order to calculate
the $\gamma$ ray flux from the dSph, the correlation between the DM density profile and velocity
dispersion at each position should be taken into account. We study such correlation and the
relevant uncertainty from kinematic observations by performing a Jeans analysis.
Using the observational results of three ultrafaint dSphs with large J-factors, including Willman 1, Reticulum II, and Triangulum II, we set constraints on the p-wave annihilation cross section in the Galaxy as an example.
\end{abstract}


\maketitle

\section{Introduction}

Astrophysical and cosmological observations suggest that the universe is composed of
$\sim4.8\%$ baryons, $\sim25.8\%$ cold dark matter (DM), and $\sim69.3\%$ dark energy \cite{2016A&A...594A...1P}.
Although DM is the dominant matter component in the universe, the nature of DM remains a mystery, and is beyond the framework of the standard model (SM). Many candidates for the
DM particle have bee proposed by the theorists, among which a kind of popular candidates
is called the weakly interacting massive particle (WIMP) \cite{1996PhR...267..195J,2000RPPh...63..793B,2005PhR...405..279B}.
The observed DM relic density can be naturally explained by the current abundance of WIMPs in the thermal
freeze-out scenario \cite{2012PhRvD..86b3506S}.

WIMPs are expected to annihilate or decay to
SM particles in the universe, and could directly or indirectly produce $\gamma$ rays.
These $\gamma$ ray signatures should be mainly generated in the regions with high DM densities, and then
can be captured by detectors. The space-borne instrument Fermi Large Area Telescope (Fermi-LAT)
is such a detector, which is sensitive to the $\gamma$ rays from range 20 MeV to
above 300 GeV \cite{2009ApJ...697.1071A}. Numerous works have been performed to study the
$\gamma$ ray signatures from DM annihilations in different astrophysical systems, such as the
galaxy clusters \cite{2010JCAP...05..025A}, galactic halo \cite{2011PhRvD..84l3005H,2012ApJ...761...91A,2012PhRvD..86h3511A,2010PhRvL.104i1302A,
2012PhRvD..86b2002A,2012JCAP...08..007W,2013PhRvD..88h2002A}, and galactic DM
substructures \cite{2012A&A...538A..93Z,2012ApJ...747..121A,2012JCAP...11..050Z}.

Among these objects, the dwarf spheroidal satellite galaxies (dSphs) of the Milky Way are
promising targets to look for the $\gamma$ ray signatures from DM annihilation, because of
their close proximity, high DM densities, low diffuse Galactic $\gamma$-ray foregrounds, and lack of
conventional astrophysical $\gamma$-ray sources \cite{1998ARA&A..36..435M,2009ApJ...696..385G}.
So far, no significant $\gamma$ ray signature over background in directions of the
known dSphs has been confirmed by Fermi-LAT observations. Therefore, the constraints on DM annihilation cross section
in dSphs have been studied in numerous works \cite{2010ApJ...712..147A,2011PhRvL.107x1302A,2017ApJ...834..110A,2011PhRvL.107x1303G,2012PhRvD..86b3528C,
2012PhRvD..86b1302G,2012APh....37...26M,2012PhRvD..86f3521B,2012JCAP...11..048H,2014PhRvD..89d2001A,
2015arXiv150302641F,2013JCAP...03..018S,2017arXiv170205266Z,2010PhRvD..82l3503E,2016PhRvD..93h3513Z,2016PhRvD..94a5018C,2017PhRvD..95l3008B,2017arXiv171100749L}.

The traditional s-wave DM annihilation process is independent of the relative velocity of DM particles.
Its thermally averaged annihilation cross section accounting for the observed DM relic density is usually quoted at a canonical value
$\langle \sigma v \rangle \sim 3\times 10^{-26}$ cm$^3$s$^{-1}$. Nevertheless, the s-wave annihilation
cross section has been constrained by the Fermi-LAT $\gamma$-ray observations from dSphs, which indicates
$\langle \sigma v \rangle $ should be smaller than the canonical value at low DM masses ($\sim \mathcal{O}(1)-\mathcal{O}(10)$ GeV)
for the annihilation channels to $\tau^+\tau^-$ or $b\bar{b}$ \cite{2014PhRvD..89d2001A}. Note that $\langle \sigma v \rangle $
generally depends on the relative velocity of DM particles. For instance, it can be expanded as
$\langle \sigma v \rangle = a+b\langle v^2 \rangle+ \mathcal{O}(v^4)$ in many models. If the p-wave annihilation is not negligible, $\langle \sigma v \rangle$ is not a constant and would be suppressed at low DM velocities. In this case, due to the small DM velocity dispersions
($\sim \mathcal{O}(1)-\mathcal{O}(10) \;\mathrm{km}\; \mathrm{s}^{-1}$) in dSphs \cite{2009ApJ...704.1274W}, $\langle \sigma v \rangle$ in dSphs is
smaller than those in the local Galactic halo and the early universe. Therefore, the constraints on the local Galactic $\langle \sigma v \rangle$ derived from dSph $\gamma$-ray observations in velocity dependent annihilation scenarios may be weaker than those for the velocity independent annihilation scenario \footnote{The constraints from other cosmological observations, such as CMB and BBN, would also change in velocity dependent annihilation scenarios (see e.g. \cite{2011PhRvD..83l3511H,2015PhLB..751..246K}).}.

In this work, we study the constraint on the velocity dependent annihilation cross section from
the Fermi-LAT dSph observations, taking into account the kinematic data. Some previous studies on the similar topic can be found in Ref. \cite{2010PhRvD..82l3503E,2016PhRvD..93h3513Z,2016PhRvD..94a5018C,2017PhRvD..95l3008B,2017arXiv171100749L,2010PhRvD..82i5007C,2011PhRvD..84g5004C}. Note that in velocity dependent annihilation scenarios, the $\gamma$-ray flux from a halo depends on the DM number density and velocity dispersion in each position. Therefore,
we should consider such correlation in the analysis rather than simply using a common velocity independent J factor. In order
to take into account this effect and the related uncertainties from astrophysical observations, we use the package CLUMPY \cite{2016CoPhC.200..336B} to perform a
Jeans analysis for the kinematic data of dSphs. By using these results, then we set constraints on the local $\langle \sigma v \rangle$ in the Galaxy
derived from dSph $\gamma$-ray observations. Although we only take p-wave annihilation as a benchmark scenario in the analysis,
the application to other velocity dependent annihilation scenarios is straightforward.

This paper is organized as follows. In sec. 2, the statistical DM density distributions for dSphs are
derived by a Jeans analysis. In sec. 3, we calculate the velocity dependent DM annihilation cross
sections with the derived DM density profiles and velocity dispersions. Then we calculate the $\gamma$-ray flux from dSphs for p-wave
annihilation, and consider the related uncertainties. By using the Fermi-LAT results, the constraints on the local Galactic $\langle \sigma v \rangle$ are set. Section 4 is our conclusion.

\section{analysis for DM density profiles of dSphs}

\subsection{Jeans analysis}

We perform a Jeans analysis to obtain DM density profiles of dSphs.
The motion of a collection of stars in a gravitational field can be described by Jeans equation, which is acquired from
the collisionless Boltzmann equation. In the case of spherical symmetry, steady-state, and negligible rotational
support, the second order Jeans equation is \cite{2008gady.book.....B,2014RvMP...86...47C}
\begin{eqnarray}\label{eq:zy1}
\frac{1}{\nu(r)}\frac{d}{dr}[\nu(r)\sigma_r^2]+2\frac{\beta_{ani}(r)\sigma_r^2}{r}=-\frac{GM(r)}{r^2},
\end{eqnarray}
where $\nu(r)$ is the three-dimensional stellar number density (3D light profile), $\sigma_r^2$ is the
radial velocity dispersion of tracers, $\beta_{ani}(r)=1-\sigma_\theta^2/\sigma_r^2$ denotes the stellar
velocity anisotropy depending on the ratio of tangential and radial velocity dispersions, and
$G$ is the gravitational constant. The enclosed mass in Eq. \ref{eq:zy1} is
\begin{eqnarray}\label{eq:zy2}
M(r)=4\pi\int_0^r\rho_{DM}(s)s^2ds,
\end{eqnarray}
where $\rho_{DM}$ is the DM density profile. The DM mass density profile rather than the total mass
density profile used here is because the contribution of the stellar component is negligible compared to
the DM halo. The solution to the three-dimensional Jeans equation is
\begin{eqnarray}\label{eq:zy3}
\nu(r)\sigma_r^2=\frac{1}{A(r)}\int_r^{\infty}A(s)\nu(s)\frac{GM(s)}{s^2}ds,
\label{JS3}
\end{eqnarray}
where $A(r)=A_{r_1}\exp[\int_{r_1}^r\frac{2}{t}\beta_{ani}(t)dt]$. The variable $r_1$ is mute and
only leads to a normalization factor, which finally cancels out in Eq. \ref{JS3} \cite{2015MNRAS.446.3002B}.
However, astrophysical observations only provide the two-dimensional projected stellar number density and velocity dispersion
along the line-of-sight. Therefore, the two-dimensional solution to the Jeans equation can be expressed as
\begin{eqnarray}\label{eq:zy4}
\sigma_p^2(R)=\frac{2}{I(R)}\int_R^{\infty}[1-\beta_{ani}(r)\frac{R^2}{r^2}]\frac{\nu(r)\sigma_r^2r}{\sqrt{r^2-R^2}}dr,
\end{eqnarray}
where $\sigma_p(R)$ is the projected velocity dispersion related to the projected radius $R$,
$I(R)$ is the projected light profile (surface brightness).

We use the Markov Chain Monte Carlo (MCMC) toolkit GreAT embedded in the CLUMPY package \cite{2016CoPhC.200..336B}
to solve the spherical Jeans equation. In Eq. \ref{eq:zy4}, the surface brightness $I(R)$ is fitted
separately with a given parametric function. The parametric DM density profile $\rho_{DM}$ and the velocity
anisotropy $\beta_{ani}$ should also be chosen. Then the Jeans analysis determine the parameters
that can best reproduce the kinematic observations. The detailed surface brightness profile, DM density profile
and velocity anisotropy are introduced in the next subsection.

When fitting to the kinematic data, the CLUMPY package
provides two types of likelihood functions, namely the binned and unbinned likelihoods. If the kinematic data
are in the form of binned velocity dispersion, we can use the binned likelihood function, while if the
data are individual velocities alone the line-of-sight, the unbinned likelihood function should be adopted.
For the binned likelihood, the standard deviation of the radii distribution should be introduced.
Therefore, the binned likelihood allows the uncertainties on both observed velocity dispersion and radius.
Compared with the binned likelihood analysis, the unbinned method does not introduce biases, and can
reduce the statistical uncertainties, especially for the ultrafaint dSphs \cite{2015MNRAS.453..849B}.
In this work, we adopt the unbinned likelihood analysis to fit the kinematic data. The form of unbinned
likelihood function is written as
\begin{eqnarray}\label{eq:zy5}
\mathcal{L}_{unbin}=\prod_{i=1}^{N_{stars}}(\frac{\exp[-\frac{1}{2}(\frac{(v_i-\bar{v})^2}{\sigma_p^2(R_i)+\Delta_{v_i}^2})]}{\sqrt{2\pi[\sigma_p^2(R_i)+\Delta_{v_i}^2]}})^{P_i},
\end{eqnarray}
where $\sigma_p(R_i)$ is the projected velocity dispersion from Eq. \ref{eq:zy4}, $v_i$ is the
velocity of the $i$-th star, $\Delta_{v_i}$ is the uncertainty of velocity measurement, and
$P_i$ is the probability of membership for each star. The line-of-sight velocity is supposed
as a Gaussian distribution, and $\bar{v}$ is the mean velocity of the distribution.

\subsection{Profile selection}

As mentioned above, the stellar surface brightness profile, DM density profile, and velocity anisotropy are essential as inputs
for CLUMPY. The stellar surface brightness is usually described by the exponential profile \cite{2009MNRAS.393L..50E},
Plummer profile \cite{1911MNRAS..71..460P}, King profile \cite{1962AJ.....67..471K}, Sersic profile \cite{1968adga.book.....S},
or Zhao-Hernquist profile \cite{1996MNRAS.278..488Z, 1990ApJ...356..359H}.
We adopt the exponential profile to fit the brightness of these dSphs, which reads
\begin{eqnarray}\label{eq:zy6}
I(R)=I_{0}\exp(-\frac{R}{r_c}),
\end{eqnarray}
where $I_{0}$ is the the normalization, and $r_{c}$ is the scale radius.

In this work, We select three dSphs, namely Willman 1 (SDSS J1049+5103), Reticulum II (DES J0335.6-5403)
and Triangulum II, in our analysis. These old and metal-poor stellar sources have almost the largest J-factors among known dSphs.

\begin{figure}[!htb]
\centering
\includegraphics[width=0.9\columnwidth, angle=0]{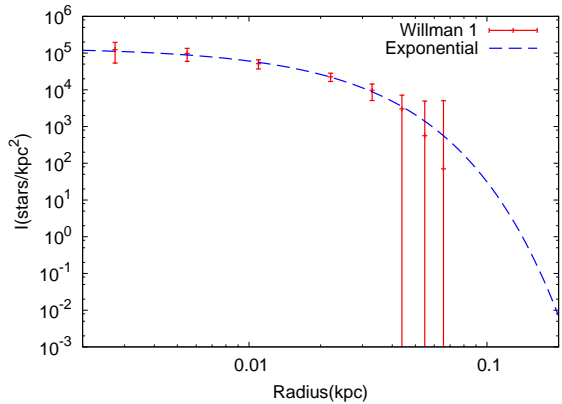}
\caption{The stellar density of Willman 1 with background subtracted. The blue dashed line
represents the best fit given by the exponential model.}
\label{fig:zy1}
\end{figure}

Willman 1 is a low-luminosity Milky Way satellite, which is located at (l, b)=($158.58^\circ, 56.78^\circ$). This source
is an old, metal-poor stellar system
at a distance of $\sim$ 38 kpc, with a half-light radius $r_h\sim$ 21 pc and an absolute
magnitude $M_V\sim$ -2.5 \cite{2005AJ....129.2692W,2006astro.ph..3486W,2007MNRAS.380..281M,2011AJ....142..128W}.
The stellar density of Willman 1 with $1\sigma$ uncertainties and background subtraction \cite{2007MNRAS.380..281M}
is shown as Fig. \ref{fig:zy1}, which is obtained from the Wide Field Camera (WFC) mounted on the Isaac Newton Telescope (INT).
We fit the stellar density with $I_{0}=1.41\times 10^5$ $\rm{stars/kpc}^2$ and $r_{c}=0.012$ kpc, shown as the
dashed line in this figure.

\begin{figure}[!htb]
\centering
\includegraphics[width=0.9\columnwidth, angle=0]{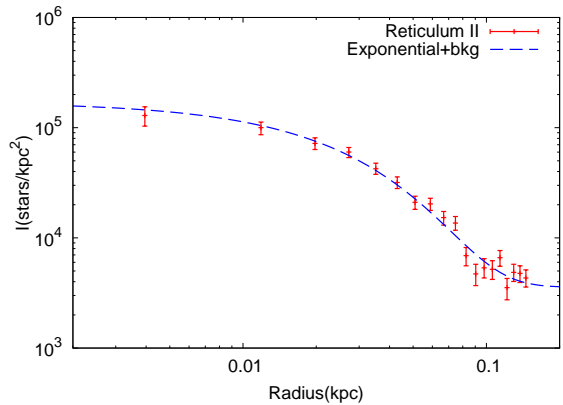}
\caption{The stellar density of Reticulum II. The blue dashed line represents the sum of the
exponential model result and background.}
\label{fig:zy2}
\end{figure}

Reticulum II was discovered from the recent optical imaging data collected
during the first year of the Dark Energy Survey (DES). It is $\sim$ 30 kpc away and is located at (l, b)=($266.30^\circ, -49.74^\circ$) with a half-light radius $r_h\sim$ 32 pc and an absolute magnitude $M_V\sim$ -2.7 \cite{2015ApJ...807...50B,2015ApJ...805..130K,2015ApJ...808..108W}.
The stellar density of Reticulum II \cite{2015ApJ...805..130K} and fitting result are shown in Fig. \ref{fig:zy2}.
We take the exponential profile plus background to fit the data, with parameters $I_{0}=1.67\times 10^5$ $\rm{stars/kpc}^2$,
$r_{c}=0.023$ kpc, and background $I_{\rm{bkg}}=3570$ $\rm{stars/kpc}^2$.

\begin{figure}[!htb]
\centering
\includegraphics[width=0.9\columnwidth, angle=0]{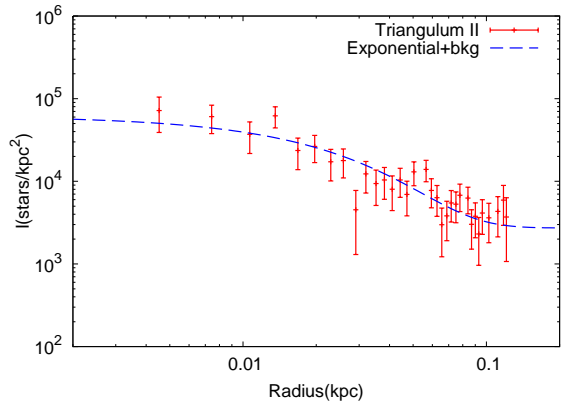}
\caption{The stellar density of Triangulum II. The blue dashed line represents the sum of the
exponential model result and background.}
\label{fig:zy3}
\end{figure}

Triangulum II was discovered from the Panoramic Survey
Telescope And Rapid Response System 3$\pi$ imaging data. It is located at (l, b)=($140.90^\circ, -23.82^\circ$),
with a heliocentric distance of $\sim$ 30 kpc, a half-light radius $r_h\sim$ 34 pc, and an absolute magnitude
$M_V\sim$ -1.8 \cite{2015ApJ...802L..18L, 2015ApJ...814L...7K, 2016ApJ...818...40M}.
The stellar density of Triangulum II \cite{2015ApJ...802L..18L} is depicted in Fig. \ref{fig:zy3}.
We also adopt the exponential model plus background to fit the data with $I_{0}=5.9\times 10^4$ $\rm{stars/kpc}^2$,
$r_{c}=0.021$ kpc, and $I_{\rm{bkg}}=2726$ $\rm{stars/kpc}^2$.

A parametric DM density profile is required in the analysis. We use the Einasto \cite{2006AJ....132.2685M}
profile in this analysis, which is written as
\begin{eqnarray}\label{eq:zy7}
\rho_{DM}(r)=\rho_{s}\exp \left\{-\frac{2}{\alpha}[(\frac{r}{r_{s}})^{\alpha}-1] \right\},
\end{eqnarray}
where $\rho_{s}$ is the normalization, $r_{s}$ is the scale radius
corresponding to the profile slope at $d \ln \rho /d \ln r=-2$, and $\alpha$ is the logarithmic slope.
These three parameters are taken to be free parameters, and should be finally acquired by the fit.

Another more flexible form of the commonly used profile is the NFW \cite{1997ApJ...490..493N} profile, which is also called
the Zhao-Hernquist \cite{1996MNRAS.278..488Z,1990ApJ...356..359H} profile. This kind of profiles can be
written as $\rho_{DM}(r)=\frac{\rho_s}{(r/r_s)^{\gamma}[1+(r/r_s)^{\alpha}]^{(\beta-\gamma)/\alpha}}$,
which relaxes the the transition slope $\alpha$, the outer slope $\beta$ and the inner slope $\gamma$.
Compared with the Einasto profile, this Zhao-Hernquist profile has more free parameters, and cause a slow
MCMC running in the numerical analysis. The choice of these two kinds of profiles would not significantly affect
the J factors for classical dSphs \cite{2015MNRAS.446.3002B}, but it will affect that for ultrafaint dSphs more or less.
We take the Enasto profile in this study.

In the Jeans analysis, the stellar velocity anisotropy profile cannot be directly derived from measured stellar velocities,
and therefore should be modeled or treated as a unknown factor. In this work we use a constant anisotropy
model written as $\beta_{ani}^{Const}(r)=\beta_0$, of which the value is freely set in the MCMC algorithm.
Another two different models can also be adopted in the analysis by the CLUMPY package. One is a simple anisotropy model
proposed as $\beta_{ani}^{Osipkov}(r)=r^2/(r^2+r_a^2)$ in Ref. \cite{1979PAZh....5...77O,1985AJ.....90.1027M},
with a single parameter of scale radius $r_a$. Another model proposed by Baes \& van Hese \cite{2007A&A...471..419B} is
a more flexible one written as $\beta_{ani}^{Baes}(r)=\frac{\beta_0+\beta_{\infty}(r/r_a)^{\eta}}{1+(r/r_a)^{\eta}}$,
where $\beta_0$ and $\beta_{\infty}$ are the anisotropies at the center and large radius respectively, $\eta$ is the
sharpness of the transition. $\beta_{ani}^{Osipkov}(r)$ is a specific form of $\beta_{ani}^{Baes}(r)$,
with the parameters $\beta_0=0$, $\beta_{\infty}=1$, and $\eta=2$. For the classical dSphs, the Baes model is recommended to
used in the fit, in order to take use of large stellar kinematic samples. However, this is not the case for the
ultrafaint dSphs. Since the available stellar kinematic samples are few, it is difficult to clearly determine the anisotropy model
for the ultrafaint dSphs considered in our study.
Therefore, we take the constant anisotropy model for simplicity.

In order to have a large statistics, we set 10 chains with $10^5$ points per chain in the MCMC analysis.
The priors of the DM density profile and the stellar velocity anisotropy models for three dSphs
are listed in Tab. \ref{tab:zy1}. The validity of these priors has been discussed in Ref. \cite{2015MNRAS.446.3002B}.

\begin{table}[!htb]
\belowcaptionskip=1em
\centering
\caption{Priors of the DM density profile (Einasto) and velocity anisotropy model (Constant)
 for three dSphs set in the MCMC analysis. $r_c$ is the light scale radius.}
\setlength\tabcolsep{1.0em}
 \begin{tabular}{lcc}
  \hline\hline\noalign{\smallskip}
   Parameter  &Unit    &Prior range  \\
  \hline\noalign{\smallskip}
  $\log_{10}(\rho_s)$   & $(M_{\odot}/kpc^3)$    & [5, 13]      \\
  $\log_{10}(r_s)$      & $(kpc)$                & [$r_c$, 1]       \\
  $\alpha$              & -                      & [0.12, 1]      \\
  \hline\noalign{\smallskip}
  $\beta_{ani}$         & -                      & [-9, 1] \\
  \noalign{\smallskip}\hline\hline
\end{tabular}
\label{tab:zy1}
\end{table}

\subsection{Kinematic data}

We use the above approach to fit the observed kinematic data of each dSph. The kinematic data are in the form of
coordinates and line-of-sight velocities with errors for individual stars. The member stars of the dSph are
often contaminated by the Milky Way foreground stars. The membership probability $P_i$ in Eq. \ref{eq:zy5} is
such a weight parameter describing whether the i-th star belongs to the dSph. Some methods have been introduced to obtain
the membership probability \cite{2009AJ....137.3109W,2009JCAP...06..014M,2007MNRAS.378..353K,2007MNRAS.377..843W}.
For the classical dSphs with large numbers of member stars, the membership probability of each star should be calculated
one by one. While for the ultrafaint dSphs, whose member stars are few, a binary classification ($P_i=0$ or $P_i=1$)
for each star based on the velocity and line-strength criteria \cite{2007ApJ...670..313S} is enough.
The data of member stars in Willman 1 and Reticulum II are taken from table 2 of Ref. \cite{2011AJ....142..128W} and
table 1 of Ref. \cite{2015ApJ...808..108W} respectively. The member stars have already been identified in their tables.
For Triangulum II, six member stars are measured by Ref. \cite{2015ApJ...814L...7K} and thirteen member stars are identified
by Ref. \cite{2016ApJ...818...40M}. In these two sets of data, five stars are overlapped.
Therefore, we adopt the fourteen member stars of Triangulum II shown in the table 2 of Ref. \cite{2016MNRAS.463.3630G}.

\subsection{DM density profile and J factor}
As mentioned above, we take the Einasto profile with three free parameters to model the DM density in the Jeans analysis.
According to the posterior probability from the fit to dSph kinematic data, a set of density profiles for each dSph are given by the CLUMPY through the MCMC algorithm. We find that the number of derived profiles for Willman 1 is the smallest among the three dSphs.
This is because that the kinematic samples of Willman 1 are larger than the other two, which
result in a more efficient convergence in the fit.

\begin{figure}[!htb]
\centering
\includegraphics[width=0.9\columnwidth, angle=0]{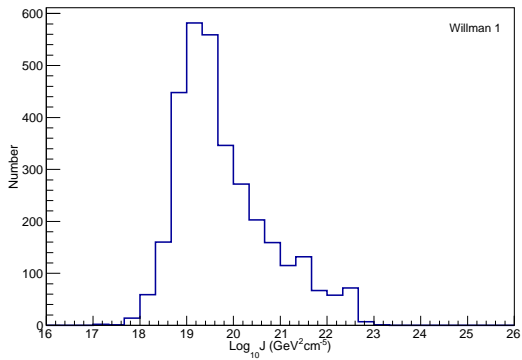}
\caption{The statistical distribution of J factors with integral angle $0.5^\circ$ for Willman 1.}
\label{fig:zy4}
\end{figure}

\begin{figure}[!htb]
\centering
\includegraphics[width=0.9\columnwidth, angle=0]{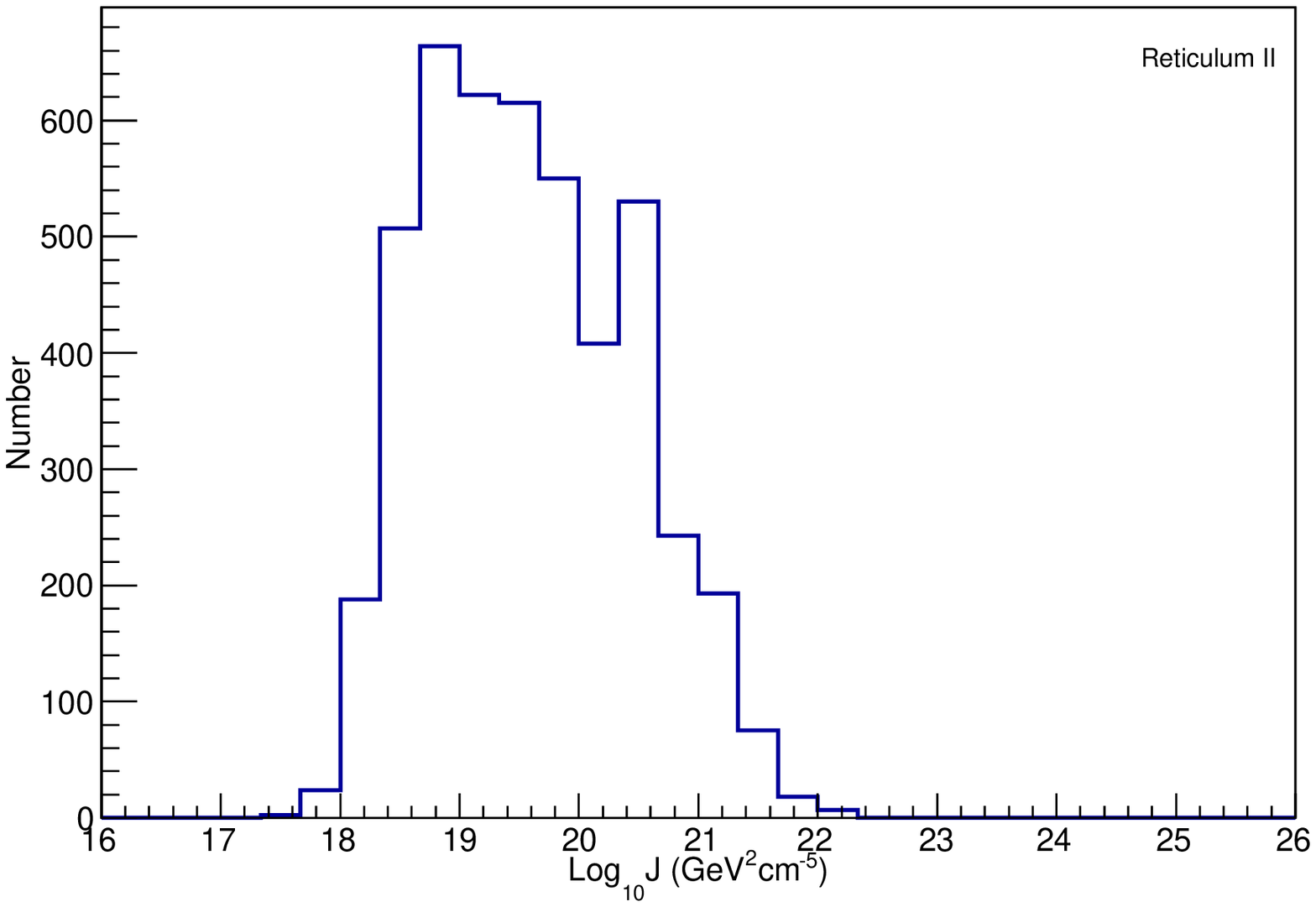}
\caption{The statistical distribution of J factors with integral angle $0.5^\circ$ for Reticulum II.}
\label{fig:zy5}
\end{figure}

\begin{figure}[!htb]
\centering
\includegraphics[width=0.9\columnwidth, angle=0]{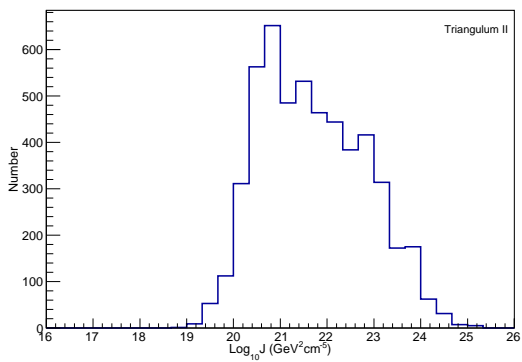}
\caption{The statistical distribution of J factors with integral angle $0.5^\circ$ for Triangulum II.}
\label{fig:zy6}
\end{figure}

A useful variable describing the dependence of the DM density distribution on the total annihilation rate is the so-called J factor.
This factor is defined as the integration of the squared DM density along the line-of-sight
over the solid angle
\begin{eqnarray}\label{eq:zy8}
J = \int_{\Delta\Omega} \int_{l.o.s.} \rho^{2}(r) dld\Omega.
\end{eqnarray}
The solid angle $\Delta \Omega$ can be expressed as $\Delta \Omega = 2\pi(1-\cos\theta)$,
where $\theta$ is the integral angle.
Then we calculate the J factor for each density profile derived from the MCMC fit, and show their
distributions in Fig. \ref{fig:zy4}-\ref{fig:zy6}.
In principle, the mean value and deviation of the J factor can be obtained from these results.

\begin{figure}[!htb]
\centering
\includegraphics[width=0.9\columnwidth, angle=0]{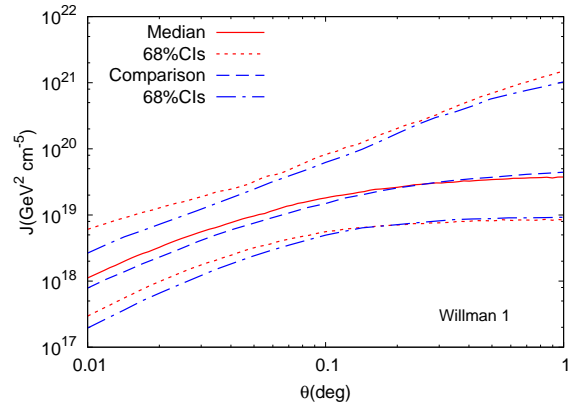}
\caption{J factor of Willman 1 as a function of integral angle. Red solid line and red dotted lines represent
the median value and $68 \%$ CIs respectively. The blue dashed line and chain lines representing the median values and $68\%$ CIs taken from Ref. \cite{2015MNRAS.453..849B} are also shown here for comparison.}
\label{fig:zy7}
\end{figure}

\begin{figure}[!htb]
\centering
\includegraphics[width=0.9\columnwidth, angle=0]{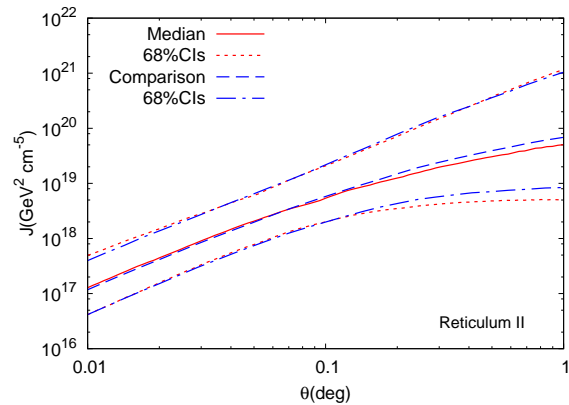}
\caption{The same as \ref{fig:zy7} but for Reticulum II. Also shown is the result from Ref. \cite{2015ApJ...808L..36B} for comparison.
}
\label{fig:zy8}
\end{figure}

\begin{figure}[!htb]
\centering
\includegraphics[width=0.9\columnwidth, angle=0]{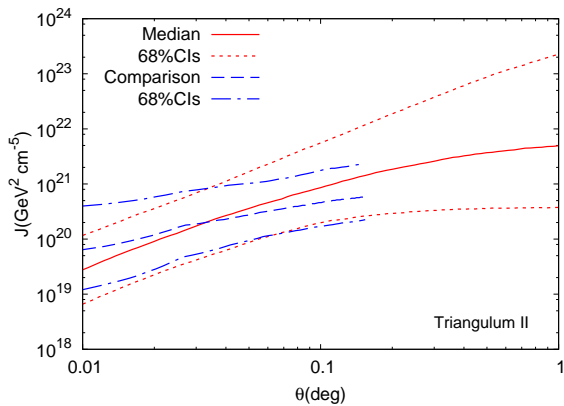}
\caption{The same as \ref{fig:zy7} but for  Triangulum II. Also shown is the result from Ref. \cite{2016MNRAS.463.3630G}.
}
\label{fig:zy9}
\end{figure}

We display the J factors of each dSph with different integral angles for three dSphs in Fig. \ref{fig:zy7}-\ref{fig:zy9}. In these figures, the red solid lines represent the median values of the J factor; the red dotted lines represent the $68\%$ confidence intervals (CIs).
In Ref. \cite{2015MNRAS.453..849B,2015ApJ...808L..36B}, the J factors of Willman 1 and Reticulum II have
been performed with different light profiles and velocity anisotropy models. We also show their median values and $68\%$ CIs
of the J factor as the blue dashed and chain lines in Fig. \ref{fig:zy7} and \ref{fig:zy8} for comparison. We find that our results
are similar with those previous results. For instance, Ref. \cite{2015MNRAS.453..849B,2015ApJ...808L..36B} obtained $\log_{10}J=19.5_{-0.6}^{+1.2}$ and $\log_{10}J=19.6_{-0.7}^{+1.0}$ for Willman 1 and Reticulum II with an integral angle of $0.5^\circ$ respectively, while we get these two values as $\log_{10}J=19.5_{-0.6}^{+1.3}$ and $\log_{10}J=19.5_{-0.8}^{+1.1}$. The
difference between the $1\sigma$ deviations may resulted from the different adopted light profiles and velocity anisotropy models.

For Triangulum II with an integration angle of $0.5^\circ$, we obtain $\log_{10}J=21.5_{-1.0}^{+1.4}$, while
the authors of Ref. \cite{2016JCAP...09..047H} used the similar approach and reported a $\log_{10}J=20.9_{-1.2}^{+1.4}$.
The difference are mostly resulted from the different samples of member stars. In Ref. \cite{2016JCAP...09..047H},
they use only six member stars measured by Ref. \cite{2015ApJ...814L...7K}. While in Ref. \cite{2016ApJ...818...40M},
thirteen member stars are identified. In these two data sets, five stars are overlapped.
Therefore, we adopt fourteen member stars of Triangulum II in the analysis.
Using the same package and fourteen member stars, the authors of Ref. \cite{2016MNRAS.463.3630G}
took the NFW DM profile, exponential light profile and constant velocity anisotropy model, and directly calculated the J factor
with the angle up to $0.15^{\circ}$. Then the J-factor with 0.5$^\circ$ of $\log_{10}J=21.03_{-0.57}^{+0.83}$ was obtained by extrapolation in Ref. \cite{2016MNRAS.463.3630G}. The difference between this result and ours are dominantly resulted from the different selections of the DM density profile.

\section{Limits on velocity dependent annihilation cross sections}

\subsection{DM annihilation cross section profile depending on relative velocities}

In this section, we calculate the $\gamma$-ray fluxes from DM annihilations in dSphs and derive the related uncertainties.
In general, the $\gamma$-ray flux is expressed as
\begin{eqnarray}\label{eq:zy9}
\Phi=\frac{1}{8\pi m_{DM}^2}\int \frac{dN_\gamma}{dE_\gamma}dE_{\gamma}\int_{\Delta\Omega} \int_{l.o.s.} \left<\sigma v\right> \rho^{2} dld\Omega,
\end{eqnarray}
where $m_{DM}$ is the mass of DM and
$\frac{dN_\gamma}{dE_\gamma}$ is the differential $\gamma$-ray spectrum from one DM pair annihilation. In this work, we only consider the contribution to $\frac{dN_\gamma}{dE_\gamma}$ from single annihilation channel given by PPPC4DM \cite{2011JCAP...03..051C,2011JCAP...03..019C}.

For velocity independent DM annihilation, Eq. \ref{eq:zy9} can be rewritten by
\begin{eqnarray}\label{eq:zy10}
\Phi=\frac{\left<\sigma v\right>\cdot J}{8\pi m_{DM}^2}\int_{E_{min}}^{E_{max}}\frac{dN_\gamma}{dE_\gamma}dE_{\gamma},
\end{eqnarray}
where J factor defined in the previous section is independent of $\left<\sigma v\right>$.
However Eq. \ref{eq:zy10} is not valid for the velocity dependent DM annihilation scenarios. Because the DM velocity distribution is not uniform in a DM halo, $\left<\sigma v\right>$ depending on the DM relative velocity changes at different positions. This means that $\left<\sigma v\right>$ cannot be simply moved from the integration in Eq. \ref{eq:zy9}.

In principle, we can assume a modeling of the velocity dependent DM annihilation cross section as
\begin{eqnarray}\label{eq:zy11}
\sigma v_{rel}=b F(v_{rel}),
\end{eqnarray}
where the function $F(v)$ should be determined by the detailed model. Then $\left<\sigma v\right>$ at any position is given by
\begin{eqnarray}\label{eq:zy12}
\left<\sigma v\right> = b\cdot \int\int F(v_{rel}) f(v_1, \textbf{r})f(v_2, \textbf{r}) dv_1^3 dv_2^3,
\end{eqnarray}
where $f(v, \textbf{r})$ is the DM velocity profile at position of $\textbf{r}$ and the relative velocity of two annihilating DM particles is
given by $v_{rel}= |\textbf{v}_1- \textbf{v}_2|$.

In this work, we assume the DM velocity distribution is a standard isotropic Maxwell-Boltzmann distribution. Although some studies have shown that deviations from the standard Maxwell-Boltzmann form are not negligible in large halos \cite{2009JCAP...01..037F,2009MNRAS.395..797V,2009MNRAS.394..641Z,2010JCAP...02..030K}, we do not consider such effects for DM in dSphs. The DM velocity dispersion is determined by the Jeans equation for DM:
\begin{eqnarray}\label{eq:zy13}
\frac{1}{\rho(r)}\frac{d}{dr}[\rho(r)\sigma_{D,r}^2]+2\frac{\beta_{D,ani}(r)\sigma_{D,r}^2}{r}=-\frac{GM(r)}{r^2},
\end{eqnarray}
where $\sigma_{D,r}^2$ and $\beta_{D,ani}(r)$ are the
radial velocity dispersion and anisotropy profile of DM respectively. With given $\beta_{D,ani}(r)$, $\sigma_{D,r}^2$ can be derived from Eq. \ref{eq:zy3} by replacing $\nu(r)$, $\sigma_r$ and $\beta_{ani}$ with $\rho(r)$, $\sigma_{D,r}$ and $\beta_{D,ani}$ respectively. As there is no enough information for the DM velocity dispersion in dSphs, we simply take an isotropic velocity dispersion with $\beta_{D,ani}=0$.

Then $\left<\sigma v\right>$ in the nonrelativistic limit is given by
\begin{eqnarray}\label{eq:zy14}
\left<\sigma v\right> &=& b \int \sqrt{\frac{2}{\pi}} \frac{1}{v_{p}^3} v_{rel}^2 e^{-\frac{v_{rel}^2}{2v_{p}^2}}F (v_{rel})dv_{rel} \nonumber \\
&\equiv & b\cdot g(r) ,
\end{eqnarray}
where $v_p^2 \sim 2 \sigma_{D,r}^2$. This means that $\left<\sigma v\right>$ is also a function of $r$. We define a
factor C to take into account the dependence of the $\gamma$ ray flux on the density profile:
\begin{eqnarray}\label{eq:zy15}
C &=& \int_{\Delta\Omega} \int_{l.o.s.} \rho^{2}(r) g(r) dld\Omega
\nonumber\\
&=&\int_0^{\theta_{max}}\int_{\theta\cdot d}^{r_{max}}\frac{4\pi r\sin\theta\rho^{2}(r)}{\sqrt{r^2-(\theta\cdot d)^2}}g(r)drd\theta,
\end{eqnarray}
where $\theta_{max}$ is the maximum integral angle, $d$ is the heliocentric distance of the dSph,
and $r_{max}$ is the maximum radius of the dSph. We take heliocentric distance as $r_{max}$ in the integration.
Finally, we can get the formula of $\gamma$ ray flux for velocity dependent annihilation scenarios, which is similar
to  Eq. \ref{eq:zy10}:
\begin{eqnarray}\label{eq:zy16}
\Phi=\frac{b \cdot C}{8\pi m_{DM}^2}\int_{E_{min}}^{E_{max}}\frac{dN_\gamma}{dE_\gamma}dE_{\gamma}.
\end{eqnarray}

\begin{figure}[!htb]
\centering
\includegraphics[width=0.9\columnwidth, angle=0]{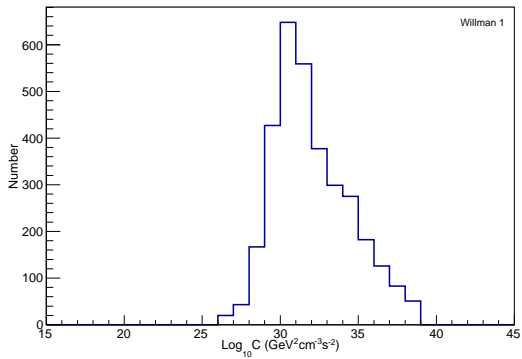}
\caption{Distribution of the C factor for Willman 1 within $0.5^\circ$.}
\label{fig:zy10}
\end{figure}

\begin{figure}[!htb]
\centering
\includegraphics[width=0.9\columnwidth, angle=0]{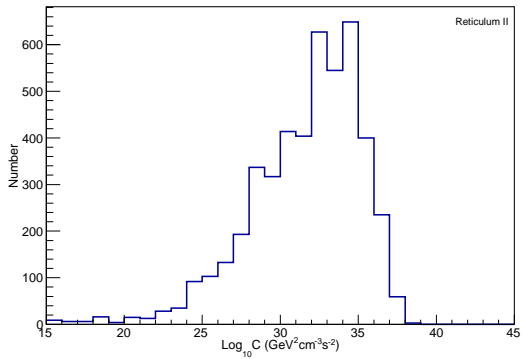}
\caption{Distribution of the C factor for Reticulum II within $0.5^\circ$. }
\label{fig:zy11}
\end{figure}

\begin{figure}[!htb]
\centering
\includegraphics[width=0.9\columnwidth, angle=0]{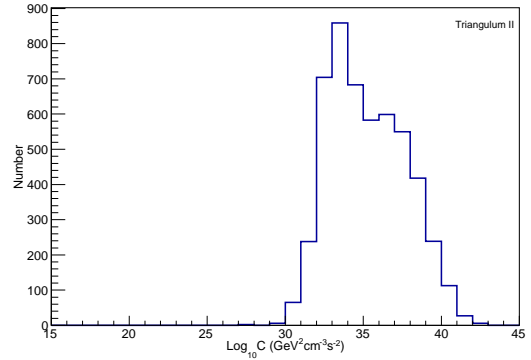}
\caption{Distribution of the C factor for Triangulum II within $0.5^\circ$.}
\label{fig:zy12}
\end{figure}

In a general WIMP scenario, the DM annihilation cross section is expanded as $\sigma v_{rel}=a+bv_{rel}^2+\mathcal O(v_{rel}^{2n})|_{n>1}$ in the nonrelativistic limit. In the following analysis, we take the pure p-wave annihilation scenario as a typical example with
\begin{eqnarray}\label{eq:zy17}
\sigma v_{rel}=bv_{rel}^2.
\end{eqnarray}
According to the density profiles derived in the previous section, we can get the corresponding distributions of C factors for dSphs.
We show such distributions for three selected dSphs in Fig. \ref{fig:zy10}-\ref{fig:zy12} and perform a gaussian fit. The mean values
and deviations of $\log_{10}C$ are obtained as $32.0\pm2.3$, $31.6\pm2.7$ and $35.4\pm2.3$ for Willman 1, Reticulum II, and Triangulum II, respectively.

\subsection{limits on local DM annihilation cross sections}

A maximum likelihood method has been developed for the Fermi-LAT $\gamma$-ray source analysis
according to the limited photon statistics and the dependence of instrumental performance
on the incident photon angle and energy. The Fermi-LAT collaboration has searched for
$\gamma$-ray emissions from dozens of dSphs using Pass 8 data \cite{2017ApJ...834..110A},
including the three sources considered in this work. In that analysis, the data set consists of six-year results
in the energy range from 500 MeV to 500 GeV, and is divided into 24 logarithmical
energy bins. The ``energy-flux-likelihood" information in the energy bin
for each dSph has been provided \footnote{\url{http://www-glast.stanford.edu/pub_data/1203/}},
and can be used for a grid scan in the likelihood analysis.
According to Eq. \ref{eq:zy16} with a parameter set of $m_{DM}$, $b$ and $C$,
the combined likelihood in all energy bins for the $j-$th dSph is estimated as
\begin{eqnarray}\label{eq:zy18}
\mathcal{L}_{j}&&=\prod_{i}\mathcal{L}_{i}(\Phi_{i}|D) \nonumber\\
&&\times \frac{e^{-[\log_{10}(C_j)-\log_{10}(C_{med,j})]^{2}/2\sigma_{j}^{2}}}{\ln(10)C_{mid,j}\sqrt{2\pi}\sigma_{j}},
\end{eqnarray}
where $i$ denotes the $i-$th energy bin, $\Phi_i$ represents the corresponding DM signature flux, and $C_{med,j}$ and $\sigma_j$ are
the obtained median value and deviation of the C factor respectively. With given $b$ and $m_{DM}$,
$C_j$ is varied to make $\mathcal{L}_j$ reach a maximum. The combined likelihood of several dSphs
is calculated by
\begin{eqnarray}\label{eq:zy19}
\mathcal{L}=\prod_{j}\mathcal{L}_{j}.
\end{eqnarray}
Then we can get upper limits on $b$ at $95\%$ C.L. by
requiring that the corresponding $\log\mathcal{L}$ has decreased by $2.71/2$ from its maximum.

\begin{figure*}[!htb]
\centering
\includegraphics[width=0.9\columnwidth, angle=0]{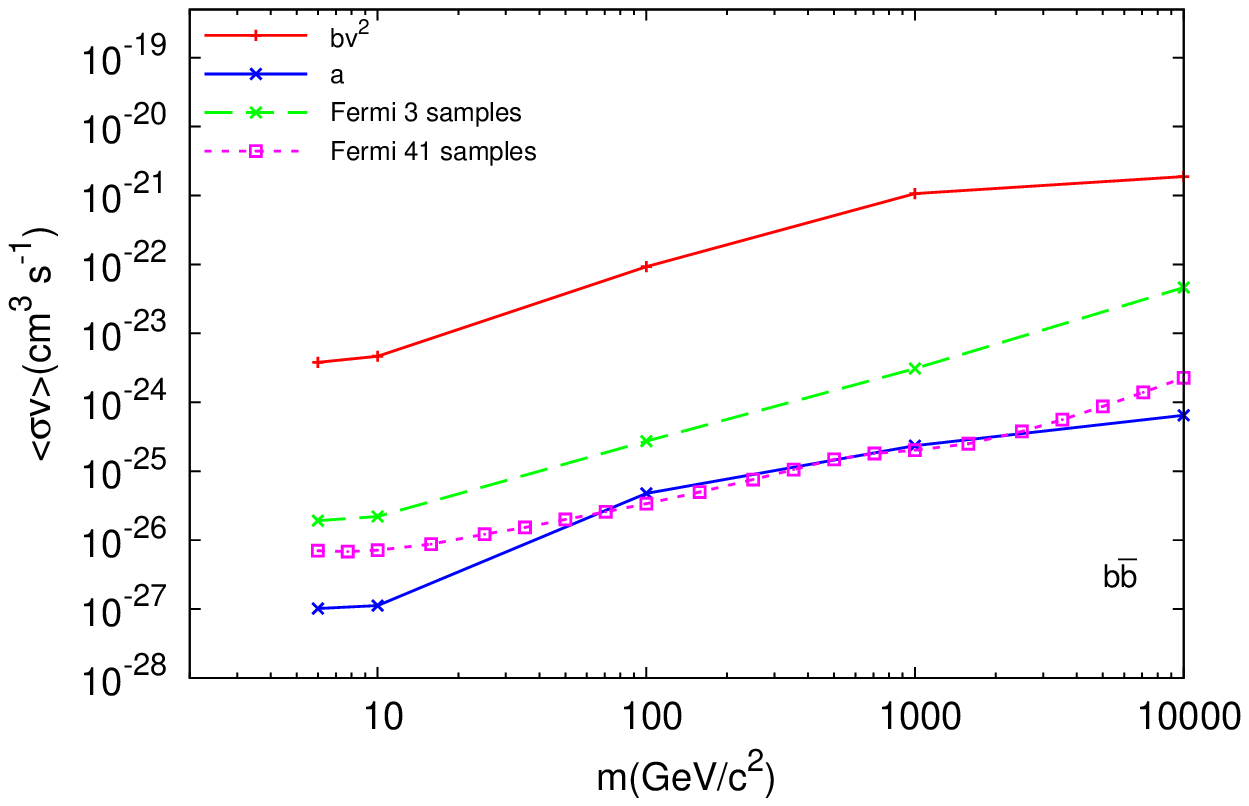}
\includegraphics[width=0.9\columnwidth, angle=0]{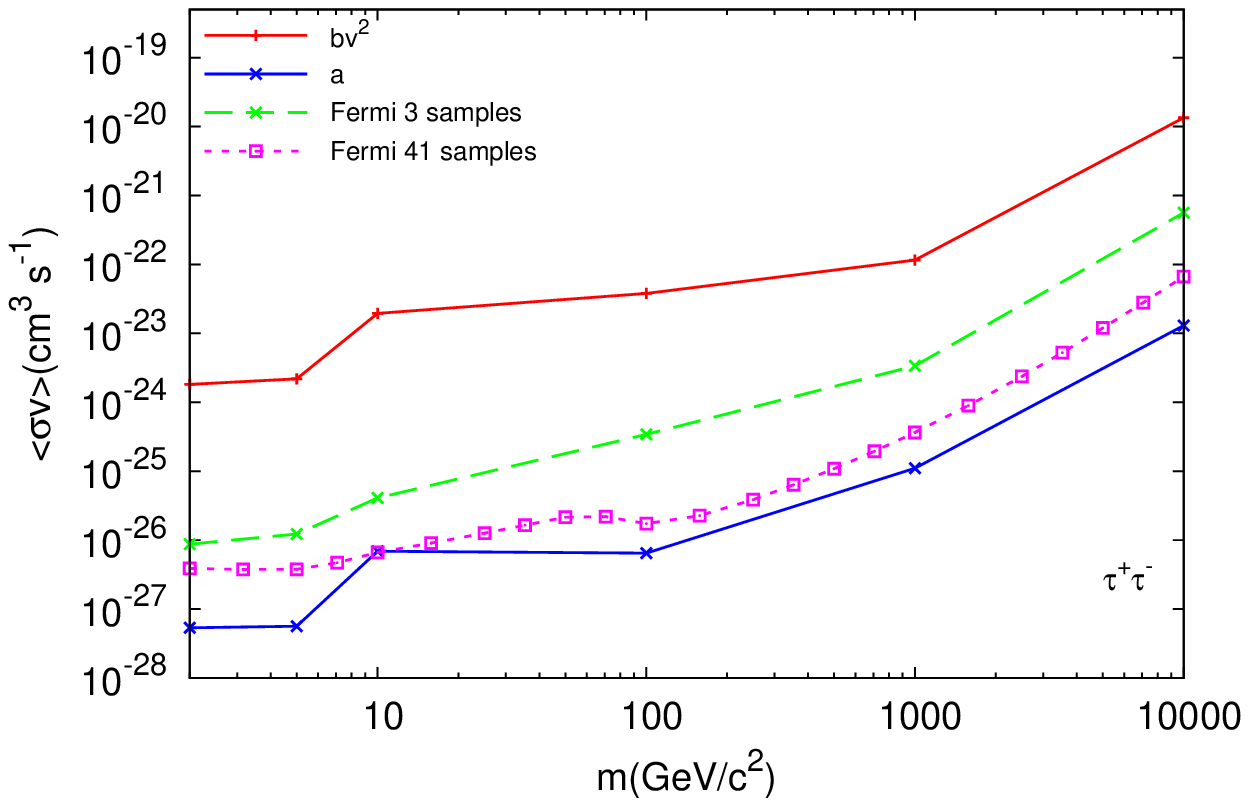}
\caption{Constraints on the DM annihilation cross section with a velocity dispersion of $\sim 270 \;\mathrm{km}\; \mathrm{s}^{-1}$ in the Galaxy at $95\%$ C.L. for the $b\bar{b}$ (left)
and $\tau^+\tau^-$ (right) channels. The red and blue solid lines represent the results for the pure p-wave annihilation
and s-wave annihilation, respectively. These results are derived from
a combined analysis of Fermi-LAT observations from Willman 1, Reticulum II and Triangulum II,
with C factors and J factors obtained in this work.
The green dashed lines are the same as blue solid lines but using the J factors as Ref. \cite{2017ApJ...834..110A}.
The pink dotted lines are given from a combined analysis for 41 dSphs in Ref. \cite{2017ApJ...834..110A}.}
\label{fig:zy13}
\end{figure*}

We translate the constraints on $b$ to those on the local DM annihilation cross section in the Milky Way with
a typical velocity dispersion of $\sim 270 \;\mathrm{km}\; \mathrm{s}^{-1}$. Such limits are useful for the comparison between results from different DM indirect detection experiments. The results for the $b\bar{b}$ and $\tau^+\tau^-$ annihilation channels are shown in Fig. \ref{fig:zy13}.
The red solid line represents the limits from three selected dSphs on local pure p-wave annihilation with
$\sigma v_{rel}\sim bv_{rel}^2$. For comparison, we also show the limits on velocity independent s-wave annihilation represented by blue solid lines, adopting the J factors obtained in our analysis. We find that they are stricter than those on p-wave annihilation by about three orders of magnitude.
For any detailed model with a given ratio of $a/b$, where $a$ and $b$ are the coefficients of the velocity expanded
cross section as mentioned above, the constraints on the DM annihilation cross section can be
obtained straightforwardly.

Finally, we briefly discuss the impact of the choice of dSph samples.
Although the J factors of the
selected three dSphs are almost the largest among the known dSphs, including more known dSphs in the analysis will
improve the constraints. The limits on the velocity independent annihilation cross section from the 41 dSphs combined analysis given by Ref. \cite{2017ApJ...834..110A} are shown in Fig. \ref{fig:zy13}. For comparison, we also show the limits for the three dSphs selected in our analysis in Fig. \ref{fig:zy13} as the green dashed lines, adopting the same J factors used in Ref. \cite{2017ApJ...834..110A} \footnote{The differences between the results represented by blue solid and green dashed lines are caused by the J factors derived from different analysis. In Ref. \cite{2017ApJ...834..110A}, the J factors of Willman 1 $\log_{10}J=18.9\pm0.6$ and Triangulum II $\log_{10}J=19.1\pm0.6$ are derived by a scaling relation between the distances and spectroscopically determined J factors of known dSphs, while the J factor of Reticulum II
$\log_{10}J=18.9\pm0.6$ is taken from Ref.~\cite{2015ApJ...808...95S}. These J factors are different from those obtained in our analysis.}. It can be seen that enlarging the samples of known dSphs would improve the limits by a factor of $\mathcal{O}(1)$ to 10.

\section{Conclusions and discussions}

In this work, we study the constraints on the velocity dependent DM annihilation cross section
from the Fermi-LAT $\gamma$-ray observations of dSphs. Since DM particles
in different astrophysical systems have different velocity dispersions, their annihilation rates may significantly
vary. Therefore, compared with ordinary analyses for the velocity independent annihilation scenario, the
constraints on the DM annihilation cross section from dSphs observations would also change.

In order to calculate the $\gamma$ ray flux from the dSph, the correlation of DM density profile and velocity
dispersion at each position should be considered. In this work, we study such correlation and
related uncertainty through Jeans analysis. We select three ultrafaint dSphs in the analysis, including Willman 1, Reticulum II, and Triangulum. These three sources have almost the largest J factors among the known dSphs. Through the Jeans analysis for kinematic data by using the package CLUMPY, a set of DM density profiles of these dSphs are given according to the posterior probability from the fit to kinematic data. We define a C factor to include the correlation between the
DM density profile and velocity dispersion. The statistical distribution of C factor is given according to the DM density profiles
derived from the Jeans analysis; the mean value and deviation can also be obtained.

We adopt the ``energy-flux-likelihood" information of every energy bin for each dSph offered
by the Fermi-LAT collaboration to set constraints on the $\gamma$ ray flux.
As an example, we set limits on the pure p-wave annihilation cross section in the Galaxy
for the $b\bar{b}$ and $\tau^+\tau^-$ channels. By using the same procedure, the analyses for other velocity dependent annihilation scenarios and annihilation channels are straightforward. Such results are useful to compare the results from
different DM indirect detection experiments. Some velocity annihilation scenarios relax the general constraints
from dSph observations, and can be used to explain some exotic results in observations. For instance, the DM explanation for the cosmic-ray $e^\pm$ excess observed by AMS-02 has almost been excluded by the Fermi-LAT dSph observations in the velocity independent
annihilation scenario, while this discrepancy can be relaxed in some velocity dependent annihilation scenarios.

In this work, we select three faint dSphs with large J factors in the Jeans analysis. Although the J factors of the three selected dSphs are almost the largest among known dSphs, there seems to exist some unclear uncertainties from the DM density density profile due to the few kinematic data of these ultrafaint dSphs. In principle, increasing the number of dSph samples would improve the final limits on the DM annihilation cross section. In the future, more dSphs may be discovered by observations. In order to search for the $\gamma$ ray signature from velocity dependent DM annihilation, careful studies on DM density profile and velocity distribution are needed.
For each dwarf galaxy, the DM density profile is correlated to the DM velocity distribution, which is the key element
for the study of the velocity dependent DM annihilation. In our analysis, we assume a standard isotropic Maxwell-Boltzmann
velocity distribution. The more information for the velocity distribution would be provided by the N-body simulation. Therefore, more kinematic data and simulation results will be helpful to construct more reliable correlation
between the DM profiles and the DM velocity distribution, and set more reliable constraints on the velocity dependent DM annihilation cross section.

\acknowledgments{We are grateful for the communications with Vincent Bonnivard and
David Maurin. This work is supported by the National Natural Science Foundation of China
under Grants Nos. 11653001, 11375202, 11475191, 11475189, by the CAS pilot B program (No. XDB23020000),
and by the National Key Program for Research and Development (No. 2016YFA0400200).}


\begin{thebibliography}{99}

\bibitem[Planck Collaboration et al.(2016)]{2016A&A...594A...1P} Planck Collaboration, Adam, R., Ade, P.~A.~R., et al.\ 2016, \aap, 594, A1

\bibitem[Jungman et al.(1996)]{1996PhR...267..195J} Jungman, G., Kamionkowski, M., \& Griest, K.\ 1996, \physrep, 267, 195

\bibitem[Bergstr{\"o}m(2000)]{2000RPPh...63..793B} Bergstr{\"o}m, L.\ 2000, Reports on Progress in Physics, 63, 793

\bibitem[Bertone et al.(2005)]{2005PhR...405..279B} Bertone, G., Hooper, D., \& Silk, J.\ 2005, \physrep, 405, 279

\bibitem[Steigman et al.(2012)]{2012PhRvD..86b3506S} Steigman, G., Dasgupta, B., \& Beacom, J.~F.\ 2012, \prd, 86, 023506

\bibitem[Atwood et al.(2009)]{2009ApJ...697.1071A} Atwood, W.~B., Abdo, A.~A., Ackermann, M., et al.\ 2009, \apj, 697, 1071

\bibitem[Ackermann et al.(2010)]{2010JCAP...05..025A} Ackermann, M., Ajello, M., Allafort, A., et al.\ 2010, \jcap, 5, 025

\bibitem[Hooper \& Linden(2011)]{2011PhRvD..84l3005H} Hooper, D., \& Linden, T.\ 2011, \prd, 84, 123005

\bibitem[Ackermann et al.(2012)]{2012ApJ...761...91A} Ackermann, M., Ajello, M., Atwood, W.~B., et al.\ 2012, \apj, 761, 91

\bibitem[Abazajian \& Kaplinghat(2012)]{2012PhRvD..86h3511A} Abazajian, K.~N., \& Kaplinghat, M.\ 2012, \prd, 86, 083511

\bibitem[Abdo et al.(2010)]{2010PhRvL.104i1302A} Abdo, A.~A., Ackermann, M., Ajello, M., et al.\ 2010, Physical Review Letters, 104, 091302

\bibitem[Ackermann et al.(2012)]{2012PhRvD..86b2002A} Ackermann, M., Ajello, M., Albert, A., et al.\ 2012, \prd, 86, 022002

\bibitem[Weniger(2012)]{2012JCAP...08..007W} Weniger, C.\ 2012, \jcap, 8, 007

\bibitem[Ackermann et al.(2013)]{2013PhRvD..88h2002A} Ackermann, M., Ajello, M., Albert, A., et al.\ 2013, \prd, 88, 082002

\bibitem[Zechlin et al.(2012)]{2012A&A...538A..93Z} Zechlin, H.-S., Fernandes, M.~V., Els{\"a}sser, D., \& Horns, D.\ 2012, \aap, 538, A93

\bibitem[Ackermann et al.(2012)]{2012ApJ...747..121A} Ackermann, M., Albert, A., Baldini, L., et al.\ 2012, \apj, 747, 121

\bibitem[Zechlin \& Horns(2012)]{2012JCAP...11..050Z} Zechlin, H.-S., \& Horns, D.\ 2012, \jcap, 11, 050

\bibitem[Mateo(1998)]{1998ARA&A..36..435M} Mateo, M.~L.\ 1998, \araa, 36, 435

\bibitem[Grcevich \& Putman(2009)]{2009ApJ...696..385G} Grcevich, J., \& Putman, M.~E.\ 2009, \apj, 696, 385


\bibitem[Abdo et al.(2010)]{2010ApJ...712..147A} Abdo, A.~A., Ackermann, M., Ajello, M., et al.\ 2010, \apj, 712, 147

\bibitem[Ackermann et al.(2011)]{2011PhRvL.107x1302A} Ackermann, M., Ajello, M., Albert, A., et al.\ 2011, Physical Review Letters, 107, 241302

\bibitem[Albert et al.(2017)]{2017ApJ...834..110A} Albert, A., Anderson, B., Bechtol, K., et al.\ 2017, \apj, 834, 110

\bibitem[Geringer-Sameth \& Koushiappas(2011)]{2011PhRvL.107x1303G} Geringer-Sameth, A., \& Koushiappas, S.~M.\ 2011, Physical Review Letters, 107, 241303

\bibitem[Cholis \& Salucci(2012)]{2012PhRvD..86b3528C} Cholis, I., \& Salucci, P.\ 2012, \prd, 86, 023528

\bibitem[Geringer-Sameth \& Koushiappas(2012)]{2012PhRvD..86b1302G} Geringer-Sameth, A., \& Koushiappas, S.~M.\ 2012, \prd, 86, 021302

\bibitem[Mazziotta et al.(2012)]{2012APh....37...26M} Mazziotta, M.~N., Loparco, F., de Palma, F., \& Giglietto, N.\ 2012, Astroparticle Physics, 37, 26

\bibitem[Baushev et al.(2012)]{2012PhRvD..86f3521B} Baushev, A.~N., Federici, S., \& Pohl, M.\ 2012, \prd, 86, 063521

\bibitem[Huang et al.(2012)]{2012JCAP...11..048H} Huang, X., Yuan, Q., Yin, P.-F., Bi, X.-J., \& Chen, X.\ 2012, \jcap, 11, 048

\bibitem[Ackermann et al.(2014)]{2014PhRvD..89d2001A} Ackermann, M., Albert, A., Anderson, B., et al.\ 2014, \prd, 89, 042001

\bibitem[Fermi-LAT Collaboration(2015)]{2015arXiv150302641F} Fermi-LAT Collaboration 2015, arXiv:1503.02641

\bibitem[Sming Tsai et al.(2013)]{2013JCAP...03..018S} Sming Tsai, Y.-L., Yuan, Q., \& Huang, X.\ 2013, \jcap, 3, 018

\bibitem[Zhao et al.(2017)]{2017arXiv170205266Z} Zhao, Y., Bi, X.-J., Yin, P.-F., \& Zhang, X.-M.\ 2017, arXiv:1702.05266

\bibitem[Essig et al.(2010)]{2010PhRvD..82l3503E} Essig, R., Sehgal, N., Strigari, L.~E., Geha, M., \& Simon, J.~D.\ 2010, \prd, 82, 123503

\bibitem[Zhao et al.(2016)]{2016PhRvD..93h3513Z} Zhao, Y., Bi, X.-J., Jia, H.-Y., Yin, P.-F., \& Zhu, F.-R.\ 2016, \prd, 93, 083513

\bibitem[Choquette et al.(2016)]{2016PhRvD..94a5018C} Choquette, J., Cline, J.~M., \& Cornell, J.~M.\ 2016, \prd, 94, 015018

\bibitem[Boddy et al.(2017)]{2017PhRvD..95l3008B} Boddy, K.~K., Kumar, J., Strigari, L.~E., \& Wang, M.-Y.\ 2017, \prd, 95, 123008

\bibitem[Lu et al.(2017)]{2017arXiv171100749L} Lu, B.-Q., Wu, Y.-L., Zhang, W.-H., \& Zhou, Y.-F.\ 2017, arXiv:1711.00749

\bibitem[Walker et al.(2009)]{2009ApJ...704.1274W} Walker, M.~G., Mateo, M., Olszewski, E.~W., et al.\ 2009, \apj, 704, 1274

\bibitem[Hisano et al.(2011)]{2011PhRvD..83l3511H} Hisano, J., Kawasaki, M., Kohri, K., et al.\ 2011, \prd, 83, 123511

\bibitem[Kawasaki et al.(2015)]{2015PhLB..751..246K} Kawasaki, M., Kohri, K., Moroi, T., \& Takaesu, Y.\ 2015, Physics Letters B, 751, 246

\bibitem[Campbell et al.(2010)]{2010PhRvD..82i5007C} Campbell, S., Dutta, B., \& Komatsu, E.\ 2010, \prd, 82, 095007

\bibitem[Campbell \& Dutta(2011)]{2011PhRvD..84g5004C} Campbell, S., \& Dutta, B.\ 2011, \prd, 84, 075004


\bibitem[Bonnivard et al.(2016)]{2016CoPhC.200..336B} Bonnivard, V., H{\"u}tten, M., Nezri, E., et al.\ 2016, Computer Physics Communications, 200, 336

\bibitem[Binney \& Tremaine(2008)]{2008gady.book.....B} Binney, J., \& Tremaine, S.\ 2008, Galactic Dynamics: Second Edition, by James Binney and Scott Tremaine.~ISBN 978-0-691-13026-2 (HB).~Published by Princeton University Press, Princeton, NJ USA, 2008.


\bibitem[Courteau et al.(2014)]{2014RvMP...86...47C} Courteau, S., Cappellari, M., de Jong, R.~S., et al.\ 2014, Reviews of Modern Physics, 86, 47

\bibitem[Bonnivard et al.(2015)]{2015MNRAS.446.3002B} Bonnivard, V., Combet, C., Maurin, D., \& Walker, M.~G.\ 2015, \mnras, 446, 3002

\bibitem[Evans et al.(2009)]{2009MNRAS.393L..50E} Evans, N.~W., An, J., \& Walker, M.~G.\ 2009, \mnras, 393, L50

\bibitem[Plummer(1911)]{1911MNRAS..71..460P} Plummer, H.~C.\ 1911, \mnras, 71, 460

\bibitem[King(1962)]{1962AJ.....67..471K} King, I.\ 1962, \aj, 67, 471

\bibitem[Sersic(1968)]{1968adga.book.....S} Sersic, J.~L.\ 1968, Cordoba, Argentina: Observatorio Astronomico, 1968

\bibitem[Zhao(1996)]{1996MNRAS.278..488Z} Zhao, H.\ 1996, \mnras, 278, 488

\bibitem[Hernquist(1990)]{1990ApJ...356..359H} Hernquist, L.\ 1990, \apj, 356, 359

\bibitem[Willman et al.(2005)]{2005AJ....129.2692W} Willman, B., Blanton, M.~R., West, A.~A., et al.\ 2005, \aj, 129, 2692

\bibitem[Willman et al.(2006)]{2006astro.ph..3486W} Willman, B., Masjedi, M., Hogg, D.~W., et al.\ 2006, arXiv:astro-ph/0603486

\bibitem[Martin et al.(2007)]{2007MNRAS.380..281M} Martin, N.~F., Ibata, R.~A., Chapman, S.~C., Irwin, M., \& Lewis, G.~F.\ 2007, \mnras, 380, 281

\bibitem[Willman et al.(2011)]{2011AJ....142..128W} Willman, B., Geha, M., Strader, J., et al.\ 2011, \aj, 142, 128

\bibitem[Bechtol et al.(2015)]{2015ApJ...807...50B} Bechtol, K., Drlica-Wagner, A., Balbinot, E., et al.\ 2015, \apj, 807, 50

\bibitem[Koposov et al.(2015)]{2015ApJ...805..130K} Koposov, S.~E., Belokurov, V., Torrealba, G., \& Evans, N.~W.\ 2015, \apj, 805, 130

\bibitem[Walker et al.(2015)]{2015ApJ...808..108W} Walker, M.~G., Mateo, M., Olszewski, E.~W., et al.\ 2015, \apj, 808, 108

\bibitem[Laevens et al.(2015)]{2015ApJ...802L..18L} Laevens, B.~P.~M., Martin, N.~F., Ibata, R.~A., et al.\ 2015, \apjl, 802, L18

\bibitem[Kirby et al.(2015)]{2015ApJ...814L...7K} Kirby, E.~N., Cohen, J.~G., Simon, J.~D., \& Guhathakurta, P.\ 2015, \apjl, 814, L7

\bibitem[Martin et al.(2016)]{2016ApJ...818...40M} Martin, N.~F., Ibata, R.~A., Collins, M.~L.~M., et al.\ 2016, \apj, 818, 40

\bibitem[Merritt et al.(2006)]{2006AJ....132.2685M} Merritt, D., Graham, A.~W., Moore, B., Diemand, J., \& Terzi{\'c}, B.\ 2006, \aj, 132, 2685

\bibitem[Navarro et al.(1997)]{1997ApJ...490..493N} Navarro, J.~F., Frenk, C.~S., \& White, S.~D.~M.\ 1997, \apj, 490, 493



\bibitem[Osipkov(1979)]{1979PAZh....5...77O} Osipkov, L.~P.\ 1979, Pisma v Astronomicheskii Zhurnal, 5, 77

\bibitem[Merritt(1985)]{1985AJ.....90.1027M} Merritt, D.\ 1985, \aj, 90, 1027

\bibitem[Baes \& van Hese(2007)]{2007A&A...471..419B} Baes, M., \& van Hese, E.\ 2007, \aap, 471, 419

\bibitem[Walker et al.(2009)]{2009AJ....137.3109W} Walker, M.~G., Mateo, M., Olszewski, E.~W., Sen, B., \& Woodroofe, M.\ 2009, \aj, 137, 3109

\bibitem[Martinez et al.(2009)]{2009JCAP...06..014M} Martinez, G.~D., Bullock, J.~S., Kaplinghat, M., Strigari, L.~E., \& Trotta, R.\ 2009, \jcap, 6, 014

\bibitem[Klimentowski et al.(2007)]{2007MNRAS.378..353K} Klimentowski, J., {\L}okas, E.~L., Kazantzidis, S., et al.\ 2007, \mnras, 378, 353

\bibitem[Wojtak \& {\L}okas(2007)]{2007MNRAS.377..843W} Wojtak, R., \& {\L}okas, E.~L.\ 2007, \mnras, 377, 843

\bibitem[Simon \& Geha(2007)]{2007ApJ...670..313S} Simon, J.~D., \& Geha, M.\ 2007, \apj, 670, 313

\bibitem[Genina \& Fairbairn(2016)]{2016MNRAS.463.3630G} Genina, A., \& Fairbairn, M.\ 2016, \mnras, 463, 3630

\bibitem[Bonnivard et al.(2015)]{2015MNRAS.453..849B} Bonnivard, V., Combet, C., Daniel, M., et al.\ 2015, \mnras, 453, 849

\bibitem[Bonnivard et al.(2015)]{2015ApJ...808L..36B} Bonnivard, V., Combet, C., Maurin, D., et al.\ 2015, \apjl, 808, L36

\bibitem[H{\"u}tten et al.(2016)]{2016JCAP...09..047H} H{\"u}tten, M., Combet, C., Maier, G., \& Maurin, D.\ 2016, \jcap, 9, 047

\bibitem[Cirelli et al.(2011)]{2011JCAP...03..051C} Cirelli, M., Corcella, G., Hektor, A., et al.\ 2011, \jcap, 3, 051

\bibitem[Ciafaloni et al.(2011)]{2011JCAP...03..019C} Ciafaloni, P., Comelli, D., Riotto, A., et al.\ 2011, \jcap, 3, 019


\bibitem[Fairbairn \& Schwetz(2009)]{2009JCAP...01..037F} Fairbairn, M., \& Schwetz, T.\ 2009, \jcap, 1, 037

\bibitem[Vogelsberger et al.(2009)]{2009MNRAS.395..797V} Vogelsberger, M., Helmi, A., Springel, V., et al.\ 2009, \mnras, 395, 797

\bibitem[Zemp et al.(2009)]{2009MNRAS.394..641Z} Zemp, M., Diemand, J., Kuhlen, M., et al.\ 2009, \mnras, 394, 641

\bibitem[Kuhlen et al.(2010)]{2010JCAP...02..030K} Kuhlen, M., Weiner, N., Diemand, J., et al.\ 2010, \jcap, 2, 030


\bibitem[Simon et al.(2015)]{2015ApJ...808...95S} Simon, J.~D., Drlica-Wagner, A., Li, T.~S., et al.\ 2015, \apj, 808, 95





\end{thebibliography}
\end{document}